# ExoMars Raman Laser Spectrometer (RLS): a tool to semi-quantify the serpentinization degree of olivine-rich rocks on Mars

Marco Veneranda[1], Guillermo Lopez-Reyes[1], Elena Pascual Sanchez[1], Agata M. Krzesińska[2], Jose Antonio Manrique-Martinez[1], Aurelio Sanz-Arranz[1], Cateline Lantz[3], Emmanuel Lalla[4], Andoni Moral[5], Jesús Medina[1], Francois Poulet[3], Henning Dypvik[2], Stephanie C. Werner[2], Jorge L. Vago[6], Fernando Rull[1]

[1] *Department of Condensed Matter Physics, Crystallography and Mineralogy, Univ. of Valladolid, Spain. Ave. Francisco Vallés, 8, Boecillo, 47151 Spain. marco.veneranda.87@gmail.com*
[2] *Department of Geosciences,* Centre for Earth Evolution and Dynamics, *University of Oslo, Norway.*
[3] *Institut d'Astrophysique Spatiale, CNRS/ Université Paris-Sud, France.*
[4] *Centre for Research in Earth and Space Science, York University, Canada.*
[5] *Department of Space Programs, Instituto Nacional de Técnica Aeroespacial (INTA), Spain.*
[6] *ESA-ESTEC, Noordwijk, Netherlands*

**Abstract**

We evaluate what will be the effectiveness of the ExoMars Raman Laser Spectrometer (RLS) to determine the degree of serpentinization of olivine-rich units on Mars. We selected terrestrial analogues of martian ultramafic rocks from the Leka Ophiolite Complex (LOC) and analyzed them with both laboratory and flight-like analytical instruments. We first studied the mineralogical composition of the samples (mostly olivine and serpentine) with state-of-the-art diffractometric (XRD) and spectroscopic (Raman, NIR) laboratory systems. We compared these results with those obtained using our RLS ExoMars Simulator. Our work shows that the RLS ExoMars Simulator successfully identified all major phases. Moreover, when emulating the automatic operating mode of the flight instrument, the RLS ExoMars simulator also detected several minor compounds (pyroxene and brucite), some of which were not observed by NIR and XRD (*e.g.* calcite). Thereafter, we produced RLS-dedicated calibration curves ($R^2$ between 0.9993 and 0.9995 with an uncertainty between ±3.0% and ±5.2% with a confidence interval of 95%) to estimate the relative content of olivine and serpentine in the samples. Our results show that RLS can be very effective to identify serpentine, a scientific target of primary importance for the potential detection of biosignatures on Mars—the main objective of the ExoMars rover mission.

## 1 Introduction

The mineralogical composition of terrestrial oceanic crust is dominated by mafic and ultramafic minerals, such as olivine ($(Mg,Fe)_2SiO_4$) and pyroxene ($XY(Si,Al)_2O_6$, in which X= Na, Ca, Mn, Fe, Mg or Li and Y= Mn, Fe, Mg, Fe, Al, Cr or Ti) (Schulte et al., 2006). Due to plate tectonic activity, sections of oceanic crust can be displaced onto land, forming the so-called ophiolite complexes (Dilek and Furnes, 2011). The different temperature and pressure conditions to which ophiolite rocks are subjected can cause a thermodynamic instability that can result in serpentinization (Schulte et al., 2006). The rate of this alteration process varies as a function of water/rock ratio and temperature (among others) and leads to the formation of serpentine (Mg,Fe-rich phyllosilicate, general formula $(Mg,Fe)_3Si_2O_5(OH)_4$)—with brucite ($Mg(OH)_2$) and magnetite ($Fe_3O_4$) as by-products,—and eventually talc ($Mg_3Si_4O_{10}(OH)_2$) in case of further aqueous alteration of the serpentinized rocks (McCollom and Bach, 2009, Huang et al., 2017). Besides the crystallization of secondary mineral products, serpentinization generates hydrogen ($H_2$), an important energy source to fuel the metabolism of chemolithoautotrophic organisms (Worman et al., 2016). Through a Fischer-Tropsch reaction, $H_2$ can also react with dissolved carbon dioxide ($CO_2$) to form short-chain hydrocarbons such as methane ($CH_4$), the basic nutrient for methanotrophic bacteria (Stamenković, 2014).

Several lines of research seeking to explore the link between ophiolite serpentinization and microbial life proliferation have been recently presented (Daae et al., 2013, Neubeck et al., 2017, Rempfert et al., 2017, Vallalar et al., 2019). The number of studies regarding rock serpentinization has increased strongly over the past years due to the potential implication of $H_2$ and $CH_4$ release for life on other celestial bodies (Müntener, 2010, Holm et al., 2015). Terrestrial ophiolites are recognized by the scientific community as appropriate Mars analogue sites recording the interaction between ultramafic-rocks and water (Blank et al., 2009, Szponar et al., 2013, Greenberger et al., 2015, Crespo-Medina et al., 2017). The analytical data gathered by Mars orbiters (Mustard et al., 2005, Pelkey et al., 2007, Amador et al., 2018, Riu et al., 2019) and rovers (Arvidson et al., 2006, 2011, Blake et al., 2013, Ehlmann et al., 2017), combined with the information obtained on Earth through the study of martian meteorite fragments (Papike et al., 2009, Taylor, 2013, Righter, 2017, Torre-Fdez et al., 2017) suggest that, as is the case for the terrestrial oceanic crust, also large areas of Mars are dominated by mafic and ultramafic rocks containing olivine as one of their major components (Mustard et al., 2005, Ody et al., 2013, Ehlmann and Edwards, 2014). More specifically, global data sets recorded by the Compact Reconnaissance Imaging Spectrometer (CRISM) onboard the Mars Reconnaissance Orbiter (MRO) have detected vast serpentine-bearing deposits that prove the past occurrence of serpentinization processes (Ehlmann et al., 2010, Michalski et al., 2017).

The importance of serpentine on Mars does not rely exclusively on the release of $H_2$ and $CH_4$ as energy sources for microbial life, but also on the potential role this phyllosilicate could play in the preservation of chemical biosignatures. The crystalline structure of phyllosilicates is characterized by parallel silicate tetrahedral sheets with interlayer sites capable of hosting microorganisms and accumulating organic molecules (Fornaro et al., 2018). However, it must be underlined that not all phyllosilicates fit this purpose since, as explained by Röling et al. (2015) (Röling et al., 2015), their elemental composition could either favor (*i.e.* by providing nutrients and protection) or inhibit (*i.e.* by having antimicrobial properties or catalyzing degradation reactions) the growth of microorganisms within their lamellar crystalline structure. For instance, laboratory assays suggested that Al-rich phyllosilicates such as kaolinite ($Al_2Si_2O_5(OH)_4$) are toxic to many microorganisms, thus potentially having a negative biological-mineral effect and, consequently, affecting biosignature accumulation (Roberts, 2004).

In this context, Fornaro et al. (2018) presented laboratory experiments in which several of the minerals detected on Mars were doped with biomarkers (adenosine monophosphate and uridine monophosphate) and exposed to UV irradiation under martian pressure and temperature conditions. The collected data show that, unlike most of the evaluated minerals, lizardite and antigorite (serpentine polytypes) do not promote any significant degradation of the evaluated biomarkers. This work provides complementary results to those gathered by Röling et al. (2015), reinforcing the idea that Fe,Mg-rich and Al-poor phyllosilicates are optimal scientific targets for the search for signs of life on Mars.

ExoMars is a collaborative program of the European Space Agency and the Russian space agency Roscosmos. The goal of the ExoMars *Rosalind Franklin* rover is to search for signs of life (Vago et al., 2017). Using its drill, the rover can collect geological samples from the martian subsurface down to a depth of 2 meters (Vago et al., 2017). The retrieved material will be imaged and delivered to the analytical laboratory of the rover (Vago et al., 2017). The sample will be crushed. The resulting particulate will be dosed onto a container where it will be inspected using the combined MicrOmega (near-infrared spectrometer) (Bibring et al., 2017), RLS (Raman Laser Spectrometer) (Rull et al., 2017) and MOMA LDMS (Mars Organic Molecule Analyzer in Laser Desorption Mass Spectrometry mode) (Arevalo et al., 2015). The ability to interrogate the exact same mineral grains using all three techniques is a unique characteristic of this mission. The results of this first analysis will be of critical importance in the selection of materials to be analyzed by MOMA in GCMS (Gas Chromatography Mass Spectrometry) configuration, whose goal is to extract, separate and identify bioorganic compounds potentially preserved in the designated mineralogical samples (Arevalo et al., 2015).

The mineralogy of the landing site selected for the ExoMars rover (Oxia Planum) stands out for the widespread occurrence of Mg,Fe-rich phyllosilicate deposits (mostly vermiculite) (Turner and Bridges, 2017). However, olivine-rich units were detected next to a putative alluvial delta located in the south-east portion of the landing ellipse. Considering the plausible serpentinization of olivine and the role that Mg,Fe-rich phyllosilicates can play in preserving chemical biosignatures on our planet, we believe it is important to evaluate the capability of MicrOmega and RLS to detect and (potentially) quantify the serpentine content of olivine-rich geological samples.

As a partial response to this need, the present work focuses on the study of serpentinized rocks sampled from the Leka Ophiolite Complex (LOC), which are part of the Planetary Terrestrial Analogue Library (PTAL) collection (Veneranda et al., 2019c).

Initially we employed X-Ray diffractometry (XRD), near-infrared spectroscopy (NIR), and Raman spectroscopy to achieve a comprehensive mineralogical characterization of the samples. Thereafter, by emulating the flight instrument's automatic operating mode (multi-point analysis on a flattened powdered sample with automatic adjustment of the RLS acquisition parameters) with our RLS ExoMars Simulator we created olivine/serpentine calibration curves and, through these, determined the serpentinization degree of LOC samples. To validate the obtained semi-quantification results, we compared the RLS ExoMars Simulator data with those provided by a state-of-the-art XRD laboratory system.

The present work seeks to provide results that may help the future ExoMars mission to select optimal scientific targets to search for molecular biosignatures on Mars.

## 2 Materials and methods

### 2.1 Analyzed materials

*2.1.1 Leka ophiolite samples*

The PTAL Project (Werner et al., 2016), funded by the European Community through the call H2020-COMPET-2015 (grant 687302), provides the scientific community with an extensive set of multi-analytical spectroscopic data gathered from over 90 terrestrial analogues resembling martian geological settings (Veneranda et al., 2019c). The list encompasses several mafic and ultramafic lithologies exposed to a broad range of alteration processes, including partially serpentinized dunite rocks from LOC. Located in the Norwegian island of Leka, LOC is the result of the uplifting of the ancient ocean crust that occurred 497 ± 2 Ma ago during the Caledonian-Appalachian mountain belt formation (Dunning and Pedersen, 1988). In recent years, several studies have focused on the detailed characterization of LOC geological units and main alteration patterns (Maaløe, 2005, Iyer et al., 2008, Plümper et al., 2012, O'Driscoll et al., 2015). As reported in the geological map provided in Figure 01, the above studies confirm that LOC is mainly composed of altered and serpentinized peridotite (dunite and harzburgite) and layered gabbro units. It also includes minor metabasalt dykes, metasediments, and pillow lavas.

Concerning altered peridotite units (the subject of this study), petrographic and geochemical analyses performed by Bjerga et al. (2015) (Bjerga et al., 2015) prove that serpentinization and carbonatation reactions produced alteration zones reaching a vertical thickness between 1 and 5 m. In detail, three main alteration-related mineral assemblages can be identified: 1) partially serpentinized peridotites, where primary minerals olivine, clinopyroxene and spinel have been replaced by serpentine, magnetite, and brucite to a limited extent; 2) strongly serpentinized peridotites, where primary minerals have been totally transformed into serpentine (80 vol.%), carbonates (15 vol.%, mainly magnesite $MgCO_3$ and dolomite $CaMg(CO_3)_2$) and magnetite; and 3) talc- and carbonate- alteration zones where previously serpentinized peridotites have been almost completely altered into talc (45-55 vol.%), carbonates (40-50 vol.%, magnesite and dolomite) and chlorite by reaction with $CO_2$-rich fluids (Bjerga et al., 2015). The petrographic study of the selected rock samples also suggests that serpentine mainly grew from olivine rather than from pyroxene (Bjerga et al., 2015). This observation is in accordance with the weathering potential index (WPI) of olivine, which has the highest value among rock-forming silicates (Delvigne et al., 1979).

Taking into account the mentioned interaction between serpentinized ultramafic rocks and chemolithoautotrophic organisms, Daae et al. (2013) (Daae et al., 2013) carried out the phylogenetic analysis of microbial communities present in altered dunite samples collected from LOC down to a depth of 1.6 m. The results confirm the detection of hydrogen-oxidizing microorganisms in the analyzed rocks, suggesting that $H_2$ released during serpentinization fuels the microbial growth of anaerobic organisms in the subsurface.

Considering the role that ultramafic rocks serpentinization could play in sustaining life and in storing bioorganic material on Mars, we chose to focus our analysis on five, partially altered dunite samples from LOC. As indicated in Figure 01, the selected analogue materials were collected at the following coordinates: N 65 06 20.4 E 11 41 18.7 (analogue LE16-0002), N 65 06 35.5 E 11 40 18.6 (LE16-0004), N 65 06 16.0 E 11 36 29.0 (LE16-0005), N 65 05 07.8 E 11 35 05.8 (LE16-0010) and N 65 05 53.5 E 11 39 54.0 (LE16-0017).

Before analysis, coarse powders (grain size up to 500 µm) were prepared by crushing the collected samples in an agate mill (McCrone Micronizer Mill) for 12 minutes to reproduce the granulometry of the crushed material that the ExoMars rover will produce. Afterwards, coarse powders (used for NIR and Raman studies) were further milled to obtain sub-samples with the grain sizes necessary to perform optimal XRD analyses (below 65 µm).

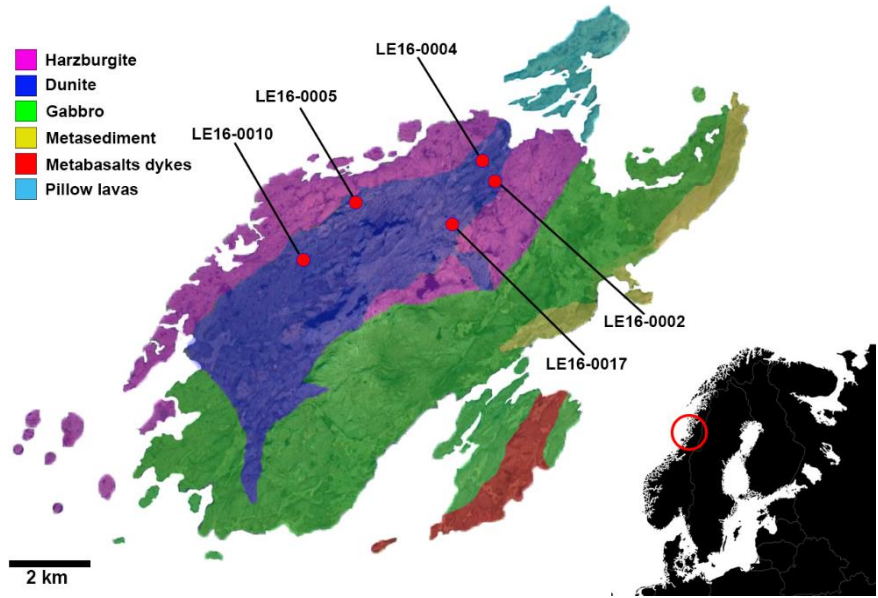

*Figure 01: Simplified geological map of the Leka Ophiolite Complex (LOC), indicating the sampling site of the analogue materials studied in this work.*

### 2.1.2 Olivine and serpentine standards

To perform a reliable semi-quantification of olivine and serpentine content of our LOC samples, we prepared an external calibration curve by mixing serpentine and olivine standards at different concentration ratios. For this purpose, we used two certified materials. According to the certificate of analysis, the serpentinite purchased from MINTEK (reference code SARM47) is mainly composed of antigorite (serpentine polytype) and contains magnetite, epidote ($Ca_2(Al,Fe)_2(SiO_4)_3(OH)$) and prehnite ($Ca_2Al(Si_3Al)O_{10}(OH)_2$) as trace minerals. On the other hand, the certified olivine purchased from the Bureau of Analysed Samples LRD (reference code SX49-12) is mostly forsterite (Mg-rich end-member of the olivine solid solution series) and contains clinochlore (($Mg,Fe^{2+})_5Al(Si_3Al)O_{10}(OH)_8$) as trace mineral.

We manually milled the certified materials using an agate mortar and, as reported in Table 01, mixed them at different concentration ratios to obtain 13 reference samples with a granulometry closely resembling the powders produced on board ExoMars.

*Table 01: Mineralogical composition of standard sample mixtures used for the external calibration procedure*

| Sample ID | Olivine (wt%) | Serpentine (wt%) |
|---|---|---|
| M0 | 0.00 | 100.00 |
| M1 | 1.10 | 98.90 |
| M5 | 5.09 | 94.91 |
| M10 | 9.99 | 90.01 |
| M25 | 24.98 | 75.02 |
| M37.5 | 37.45 | 62.55 |
| M50 | 50.10 | 49.90 |
| M62.5 | 63.34 | 37.66 |
| M75 | 75.10 | 24.90 |
| M90 | 89.94 | 10.06 |
| M95 | 94.89 | 5.11 |
| M99 | 98.90 | 1.10 |
| **M100** | 100.00 | 0.00 |

We studied the resulting samples following the analytical procedure described in sections 2.2.2 and 2.2.3.

**2.2 Analytical instruments and techniques**

The analytical procedure defined for this work includes the use of both laboratory and flight-representative analytical instruments.

*2.2.1 Laboratory instruments*

We performed X-ray analyses with a D8 Advance diffractometer (Bruker). The instrument is equipped with a Cu X-ray tube (wavelength 1.54 Å) and a LynxEye detector. We placed fine-powdered rocks in the XRD sample holder system and analyzed them in a step scan mode by setting a scan range between 2 and 65° 2θ, with a step increment in 2θ of 0.01 and a count time of 0.3 seconds per step. We used the BRUKER DIFFRAC.EVA software and the Powder Diffraction File PDF-2 2002 mineral database (ICDD) for the mineral identification of the analyzed analogues. Afterwards, we carried out a semi-quantitative estimation of the detected phases through the Profex software by using the Rietveld refinement program BGMN (Doebelin and Kleeberg, 2015).

Being the laboratory spectrometer ensuring the higher spectral resolution (4 $cm^{-1}$) the microRaman system was selected for the Raman characterization of coarse-powdered samples. We assembled this instrument in our laboratory. It is composed of the following commercial components: a Research Electro-Optics LSRP-3501 excitation laser (Helium-Neon) emitting at 633 nm, a Kaiser Optical Systems Inc. (KOSI) a HFPH Raman probe, a KOSI Holospec1.8i spectrometer and an Andor DV420A-OE-130 CCD detector. The instrument is coupled to a Nikon Eclipse E600 microscope and equipped with interchangeable long working distance (WD) objectives of 5x, 10x, 20x, 50x and 100x. Depending on the mineralogical heterogeneity of the sample under study, the operator would select between 20 and 39 points of interest through the camera coupled to the microscope. We used the Hologram 4.0 software to acquire Raman spectra from the selected mineral grains by manually setting a range of analysis between 130 and 3780 $cm^{-1}$ (average spectral resolution of 4 $cm^{-1}$), an integration time from 5 to 60 s, and a number of scans between 5 and 20. We performed the mineral phase identification by comparing our results with Raman reference spectra of pure mineral standards.

We also employed a Fourier Transform spectrometer (PerkinElmer Spectrum 100N FTNIR) for complementary near-infrared analyses. We acquired NIR spectra of coarse powdered samples by setting a range of analysis between 0.8 and 4.2 µm, a spectral resolution of 4$cm^{-1}$, and a spectral sampling of 0.35 nm. A collecting spot size of about 1-mm-diameter ensured a representative bulk characterization of all sample constituents. To calibrate the sample reflectance spectrum, we obtained reference spectra using Infragold and Spectralon 99% standards (Labsphere), while OH signatures due to ambient air (around 1.4, 1.9, and 2.7 µm) were automatically corrected by the instrument. We carried out the phase identification by comparison with NIR reference spectra of pure mineral standards. Afterwards, we evaluated the alteration degree of LOC samples by following the methodology previously developed for Mars surface remote study (Poulet et al., 2009). Specifically, sample spectra were reproduced by simulating the Light-Matter Interaction of a mineral mixture from a radiative transfer modeling based on the Shkuratov theory (Shkuratov et al., 1999). As the optical constants are not known for all minerals, we used proxies to mimic major mineral families.

*2.2.2 Flight-derived analytical instrument*

Besides the use of laboratory instruments, we carried out further Raman analyses with the RLS ExoMars Simulator (Lopez-Reyes et al., 2013). As mentioned previously, this instrument is considered the most reliable tool to simulate the automatic operational mode of the RLS onboard the *Rosalind Franklin* rover and reliably predict its potential scientific outcome (Veneranda et al., 2019c, 2019a, 2019b). Developed by the UVa-CSIC-CAB Associated Unit ERICA (Spain) the instrument setup includes a continuous 532 nm (green) excitation laser, a high resolution TE Cooled CCD Array spectrometer, an optical head with a long WD objective of 50x, and is coupled to an emulator of the Sample Preparation and Distribution System (SPDS) of the ExoMars rover. Range of analysis (70-4200 cm$^{-1}$), working distance (≈ 15 mm), laser power output (20 mW), spectral resolution (6-10 cm$^{-1}$) and spot of analysis (≈ 50 µm) all closely resembling those of the RLS flight instrument. Furthermore, the RLS ExoMars Simulator integrates the same algorithms that will be employed by RLS to autonomously operate on Mars, such as sequential analyses, autofocus, optimization of the signal to noise ratio by automatically selecting the best acquisition parameters(Lopez-Reyes and Rull Pérez, 2017). We collected multiple spectra datasets from each powdered sample emulating the operational constraints of the RLS instrument on Mars by automatically adjusting the abovementioned parameters. We acquired the data using a custom-developed software based on LabVIEW 2013 (National Instruments).

*2.2.3 Data analysis for quantification of results from the RLS ExoMars Simulator*

The *Rosalind Franklin* rover includes a drill having a maximum depth reach of two meters. After collection, a sample will be imaged and then crushed, placed on a moving carrousel, and flattened to be sequentially analyzed by the instruments of the Analytical Laboratory Drawer (ALD) of the rover, which include the RLS instrument. The rover carrousel will move to present the sample to the Raman instrument, which will analyze between 20 and 39 points, depending on the available time and resources during the sample analysis.

Profiting from the multi-point analysis performed by RLS on the sample, it is possible to establish analytical procedures to perform a semi-quantification of the mineral abundance on sample mixtures. As part the work reported here, we have prepared calibration curves for the quantification of olivine and serpentine that benefit from the acquisition of multiple points. In order to obtain (some degree of) statistical and uncertainty data for the calibration curves, we obtained five 39-point lines for each mixture proportion.

The calibration curves were calculated following the method by Lopez-Reyes et al. (Lopez-Reyes et al., 2013), in which the ratio between the intensity of non-overlapping peaks of the endmembers with respect to their total intensity are averaged for all the spectra along a n-point line:

$$r_{oli} = \frac{1}{n} \cdot \sum_n \frac{I_{oli}}{I_{oli} + I_{serp}} ; \quad r_{serp} = \frac{1}{n} \cdot \sum_n \frac{I_{serp}}{I_{oli} + I_{serp}} ;$$

where r is the estimated proportion, n is the number of spots per line and *I* is the peak intensity of the spectrum at a given spot.

The calibration curve uncertainty estimation was calculated using the standard deviation (σ) between different lines with a confidence interval of 95% (*i.e.* representing the uncertainty with a value of ±2σ). In order to avoid overlapping with minor elements found on LOC samples, the selected peak positions used for the calculation of the index are the ~230 cm$^{-1}$ for serpentine and the ~855 cm$^{-1}$ for olivine. We carried out all calculations using MATLAB R2019a.

## 3 Results

### 3.1 Characterization of Leka ophiolite samples by means of laboratory systems

*3.1.1 XRD analysis*

All LOC samples represent ultramafic rocks, composed of olivine and augite that were in various extend altered to serpentine minerals and minor Fe-oxides and Mg-hydroxides (such as brucite). We inferred the main rock-forming minerals from XRD patterns based on the presence of the following diagnostic peaks: olivine based on 17.4 and 22.8 2θ values (best match with forsterite PDF 31-0795), augite based on 27.5, 29.8 and 30.3 2θ values (fit of PDF 24-0203), serpentine at 12.2 , 24.4 and 24.6 2θ values (fit by mixture of PDF 50-1606 and PDF 02-0100) and brucite with diagnostic peak at 18.7 2θ (fit by PDF 07-0239). Additionally, all samples revealed weak peaks suggesting the presence of minor amounts of Fe-oxides (either magnetite, hematite or ilmenite).

Serpentinization degree of all samples is high but varied. A clear, reverse correlation is seen between abundance of serpentine and olivine and augite. The sample of LE16-0004 analogue is composed of 50% (by volume) serpentine, accompanied by 30% of olivine and 10% of augite. Samples labeled as LE16-0005, LE16-0010 and LE16-0017 contain 70-75 % of serpentine, 1-5 % of augite and 15-20% of olivine. In contrast, the most serpentinized sample, LE16-0002 is composed of 90% of serpentine, contains no olivine and 10% of augite.

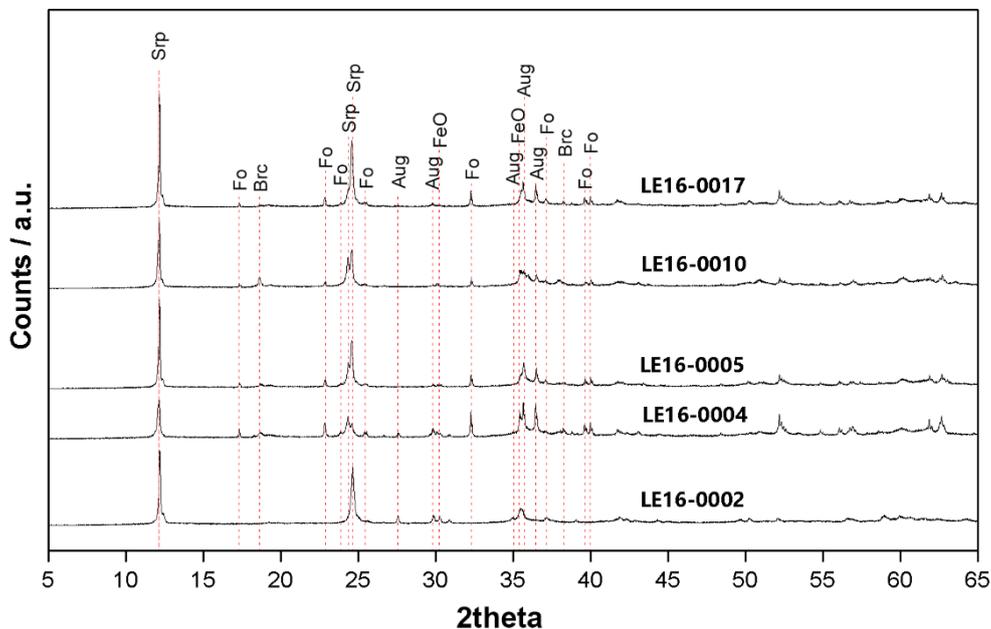

*Figure 02: XRD results gathered from Leka samples. Srp=serpentine; Fo=Forsterite; Brc=brucite; Aug=Augite; FeO=Fe-oxides;*

*3.1.2 NIR analysis*

NIR spectra from samples LE16-0002, LE16-0004, LE16-0005, LE16-0010 and LE16-0017 displayed very similar vibrational features. As can be seen in Figure 03, the five analogues provided a wide range of signals between 1.3 and 3.0 µm, confirming the presence of serpentine as main mineral phase. Regarding sample LE16-0002, the main bands at 1.38 and 2.72 µm slightly shift towards higher wavelengths (1.385 and 2.73 µm), which is indicative of a different metal abundance. Furthermore, the NIR spectra from sample LE16-0002 also displayed a clear signal at 0.92 µm, which is consistent with the hematite. Additional silicate signatures centred near 1.0 (band 1) and 2.25 µm (band 2) were detected in all samples. These bands can be attributed to both olivine (band 1) and pyroxene (bands 1 and 2).

Following the methodology developed for the interpretation of Mars data collected from orbit (Lopez-Reyes and Rull Pérez, 2017), we modelled each spectrum by simulating the Light-Matter Interaction of a mineral mixture from a radiative transfer modelling (Poulet et al., 2009). Pure lizardite, forsterite, diopside, pigeonite and augite were used as mineral proxies. The obtained models helped defining the degradation degree of Leka samples. In this regards, samples LE16-0005 and LE16-0017 display a similar alteration degree, whose are lower than LE16-0002 and higher than samples LE16-0004 and LE16-0010.

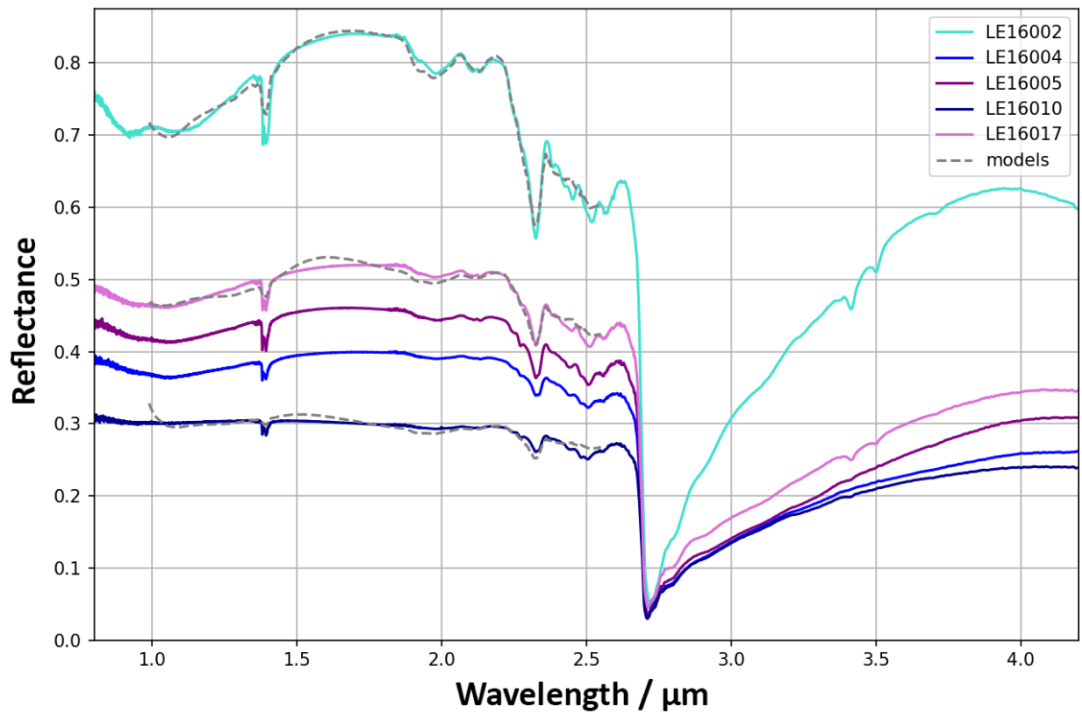

*Figure 03: experimental and modelled NIR spectra obtained from sample LE16-0002, LE16-0004, LE16-0005, LE16-0010 and LE16-0017.*

*3.1.3 Raman analysis*

Starting from sample LE16-0002, we collected 20 spectra by visually selecting the most interesting spots of analysis (minerals grains showing different color, shape and brightness). From many of the selected crystals, the typical vibrational profile of serpentine was obtained. In all serpentine spectra, we constantly detected a peak of medium intensity at 1045 cm$^{-1}$ together with main signals at 229, 379, 669, 3670 and 3693 cm$^{-1}$. According to Rinaudo et al. (Rinaudo et al., 2003), the detection of this peak indicates the presence of antigorite, a serpentine polytype characterized by an alternating-wave crystalline structure, where the orientation of octahedral layer is constant while tetrahedral one is periodically switching.

In addition to antigorite, two Raman spectra also displayed the characteristic peaks of augite (229, 324, 390, 666 and 1012 cm$^{-1}$), a pyroxene mineral including calcium, iron and magnesium ions in its crystalline structure (($Ca,Mg,Fe)_2(Si,Al)_2O_6$). One of the collected spectra also displayed weak signals at 221, 289, 404, 606, 656 and 1316 cm$^{-1}$, suggesting the presence of hematite ($Fe_2O_3$).

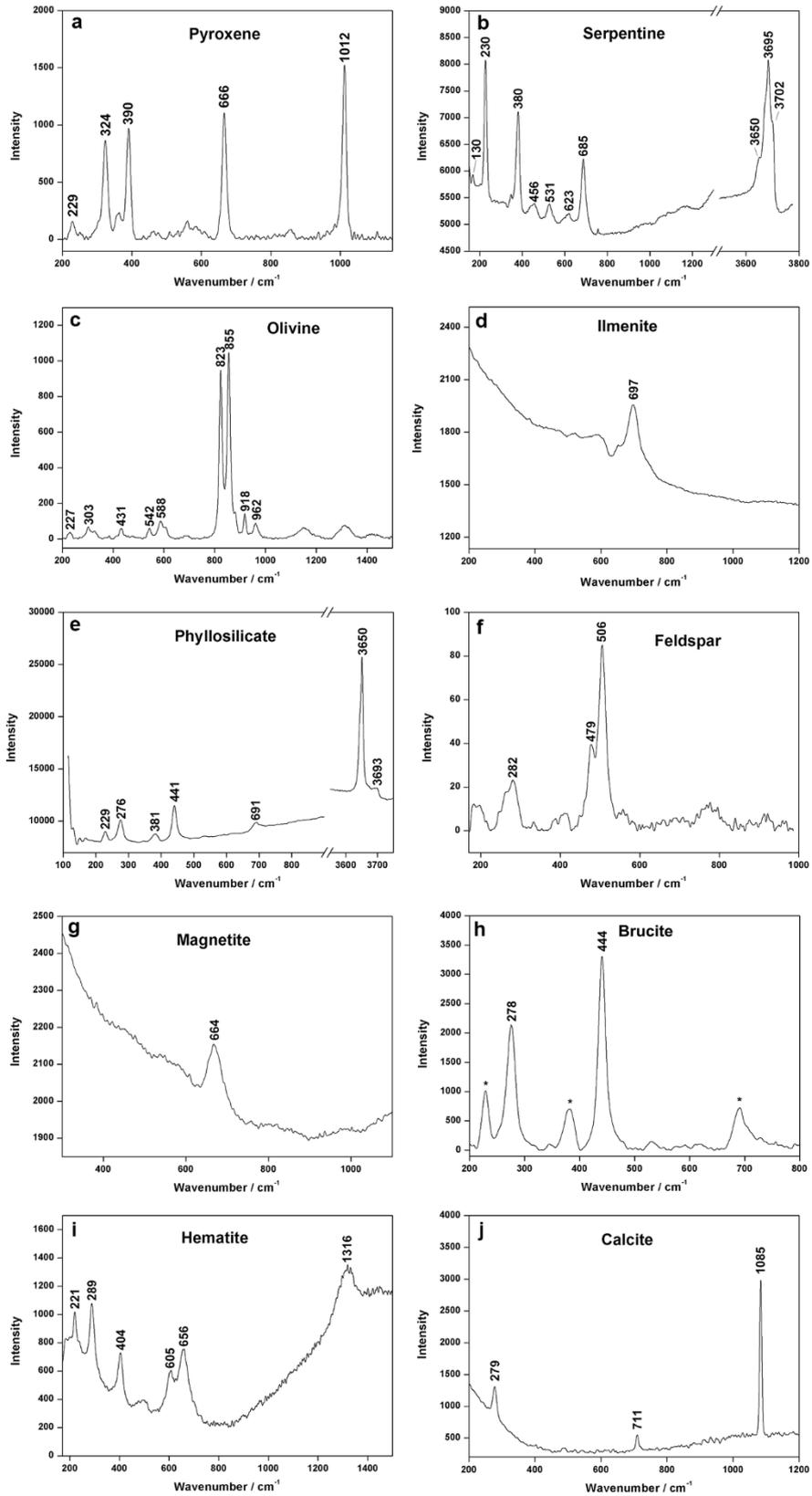

*Figure 04: representative Raman spectra collected from Leka samples by means of the microRaman system. Peaks labelled with an asterisk originate from additional compounds.*

The Raman study of samples LE16-0004 enabled the detection of pyroxene (augite) and serpentine, together with a considerable amount of olivine. With regards to serpentine, we identified the lizardite polytype (main peaks at 132, 229, 378, 668, 3674 and 3694 cm$^{-1}$) together with antigorite. As explained elsewhere, the crystalline structure of this serpentine can be distinguished by its planar structure composed of parallel and alternated octahedral and tetrahedral layers (Rinaudo et al., 2003). Regarding olivine spectra, the two characteristic peaks were detected in a wavelength range that goes from 822 to 823 cm$^{-1}$ and from 854 and 856 cm$^{-1}$ respectively (Figure 04c). As proved in previous works, the wavelength position of the doublet can be used to determine the iron/magnesium ratio of the crystal under study, i.e. content of forsterite ($Mg_2SiO_4$) and fayalite ($Fe_2SiO_4$), the end-members of the olivine solid solution (Torre-Fdez et al., 2017). We applied the calibration equation proposed by Mouri and Enami (Mouri and Enami, 2008), measuring an average composition between $Fo_{87}Fa_{13}$ and $Fo_{92}Fa_{08}$.

The Raman study of sample LE16-0005 disclosed a mineral composition dominated by serpentine (mostly antigorite), together with minor amounts of olivine ($Fo_{87}Fa_{13}$). Furthermore, Raman results proved the additional presence of ilmenite (Figure 04d, main band at 697 cm$^{-1}$), a titanium oxide ($FeTiO_3$) that was not detected by XRD or NIR.

The set of spectra gathered from sample LE16-0010 suggests a strong serpentinization degree. Indeed, antigorite and lizardite were both found as major phases, while olivine was detected in some spots only. As can be observed in Figure 04e, we collected an additional spectrum showing several peaks from weak to medium intensity at 229, 276, 381, 441 and 691 cm$^{-1}$, together with very strong signals in the OH stretching region (3650 and 3694 cm$^{-1}$). Comparing with similar spectra available in bibliography, the vibrational profile of this Raman spectra could be attributed to a mineral belonging to the phyllosilicate group (Wang et al., 2015). According to XRD semi-quantitative results, LE-16-0010 is the sample having the highest content of Mg-hydroxide (7%). This interpretation was confirmed by Raman analysis, since the main peaks of brucite ($Mg(OH)_2$) at 278 and 444cm$^{-1}$ were detected in a few of the analyzed spots (Figure 04h).

Beside the partial alteration of olivine (between $Fo_{87}Fa_{13}$ and $Fo_{92}Fa_{08}$) into serpentine (lizardite), we identified iron oxides (magnetite and hematite) on sample LE16-0017 (Figure 04i). Furthermore, as shown in Figure 04j, the detection of a strong peak at 1085 cm$^{-1}$, together with weaker signals at 279 and 711 cm$^{-1}$ confirmed the presence of carbonates (calcite, $CaCO_3$) that, as previously explained, could be interpreted as an alteration mineral crystallized during serpentinization (Bjerga et al., 2015).

### 3.2 Characterization of Leka ophiolite samples by means of space-derived instrumentation (RLS ExoMars Simulator)

We used the ExoMars RLS Simulator to extrapolate solid inferences about the scientific capabilities of the RLS instrument that will soon be deployed on Mars. By following the analytical procedure established for the ESA rover mission, we employed the RLS ExoMars Simulator to automatically perform raster analysis of 39 spots from each powdered analogue sample (Bost et al., 2015). To avoid possible overlaps between adjacent spots (diameter of 50 μm), analyses were carried out by moving the sample along the X-axis at constant intervals of 150 μm. Each spectrum was automatically acquired using RLS-like algorithms, while data processing and interpretation was done by means of the Instrument Data Analysis Tool (IDAT) and SpectPro. IDAT/SpectPro software was developed by the RLS team to receive, decodify, calibrate and verify the telemetries generated by the RLS instrument on Mars (Guillermo Lopez-Reyes et al., 2018), but also to provide analytical tools for spectral analysis.

Most of the spectra obtained from sample LE16-0002 showed the vibrational profile of serpentine, while pyroxene was only sporadically detected. It must be mentioned that the RLS ExoMars Simulator was able to clearly detect main and secondary peaks of both compounds. By comparison with reference patterns, the vibrational profile of pyroxene fit with augite (Figure 06a), while most of the serpentine spectra displayed the characteristic peak of antigorite at 1045cm$^{-1}$ (Figure 06a). The whole set of analytical data presented in this study indicates that, unlike pyroxene, olivine crystals from sample LE16-0002 were almost totally altered into serpentine. This fact agrees with petrographic analyses of Leka peridotite samples presented elsewhere (Bjerga et al., 2015), suggesting that serpentinization process mainly affects olivine grains rather than pyroxene ones.

Sample LE16-0004 differs from LE16-0002 by a lower degree of alteration. Indeed, not-altered olivine grains were clearly detected together with pyroxene (augite) and serpentine. In this case, no Raman peaks between 1000 and 1100 cm$^{-1}$ were observed on serpentine spectra, confirming the presence of lizardite polytype. As can be observed on Figure 06, another difference between lizardite and antigorite spectra can be observed in the range of wavelength between 3600 and 3750 cm$^{-1}$, were the vibrational modes of the -OH group are detected. All antigorite spectra gathered from the use of the RLS ExoMars Simulator displayed Raman signals at 3674 and 3694 cm$^{-1}$, while lizardite showed a combination of three bands at 3640, 3689 and 3706 cm$^{-1}$. We observed this difference in all the analyzed samples. Thus, the vibrational profile of hydration waters must be seen as an additional parameter to consider for the proper discrimination between serpentine polytypes.

Concerning olivine spectra, we made use of the same calibration line described in section 3.1.3 to estimate the fosterite/fayalite ratio. Considering that the main double peak was constantly detected at 822 and 854 cm$^{-1}$ (Figure 05b), the Fo/Fa value calculated from RLS ExoMars Simulator spectra coincides with that obtained from laboratory Raman data (Fo$_{87}$Fa$_{13}$).

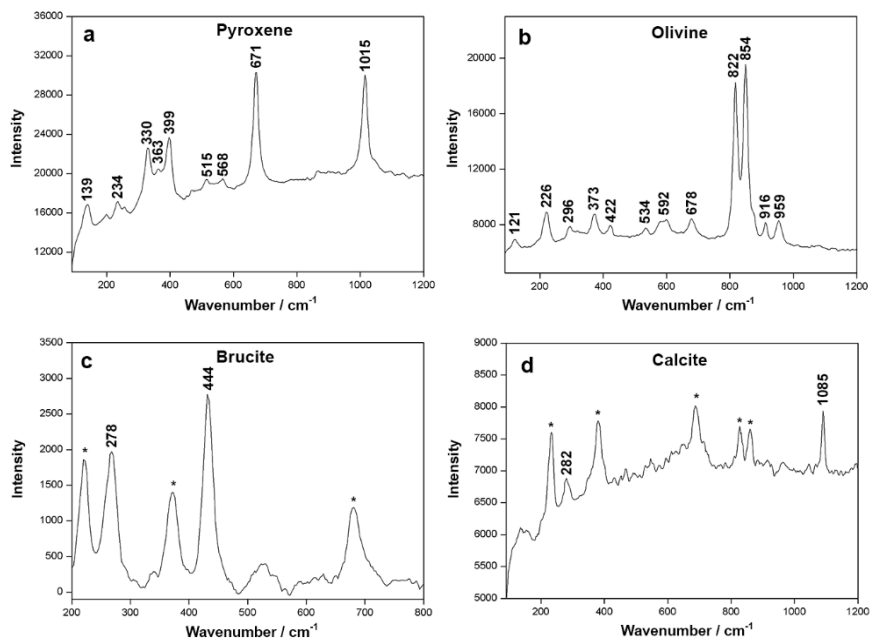

*Figure 05: representative Raman spectra collected from Leka samples by means of the RLS ExoMars Simulator system. Peaks labelled with an asterisk originate from additional compounds.*

From the Raman analysis of samples LE16-0005 and LE16-0010, we mainly detect olivine and serpentine spectra. According to the detailed analysis of olivine spectra, the average composition is $Fo_{92}Fa_{08}$ in both analogues. With regards to serpentine, sample LE16-0005 is mostly composed of antigorite while, in the case of sample LE16-0010, the ratio between antigorite and lizardite is close to 60/40. As can be seen in Figure 06c, RLS ExoMars Simulator analysis also enabled the detection of brucite main peaks at 278 and 444 $cm^{-1}$, thus confirming the presence of Mg-hydroxide on sample LE16-0010.

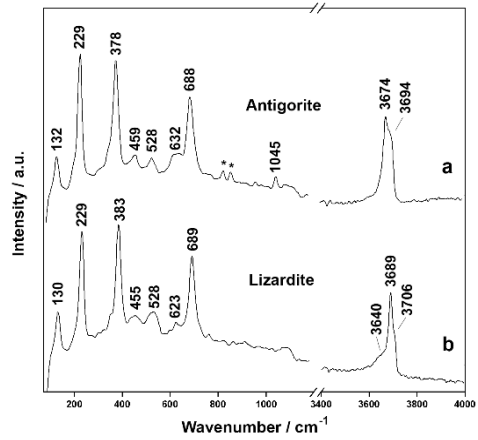

*Figure 06: Antigorite and lizardite spectra (serpentine polytypes) obtained from Leka samples by means of the RLS ExoMars Simulator system.*

We clearly detected olivine ($Fo_{92}Fa_{08}$), pyroxene (augite) and serpentine (mostly lizardite) in sample LE16-0017. In addition to those, RLS ExoMars Simulator data confirmed the identification of calcite (Figure 05d). Considering that this compound was not found by XRD and NIR, this result demonstrates that, by replicating the operational mode established for the ESA rover mission, the RLS ExoMars Simulator is able to effectively detect minor mineral phases.

**3.3 Calibration curves for olivine and serpentine semi-quantification (RLS ExoMars Simulator)**

The five 39-point lines obtained by the RLS ExoMars Simulator of the samples prepared with controlled proportions of olivine and serpentine were used to generate calibration curves following the intensity ratio method described in the Materials and Methods section.

As described by Lopez-Reyes et al. (Lopez-Reyes et al., 2013), the quantification index for each line is calculated as the average of the intensity ratio calculated for each spectrum, so the number of analyzed spots per line can potentially influence the results. We evaluate this dependence by calculating two different calibration curves: using a subset of the total number of acquired spectra (20 points por line, which is the minimum number of analysis the RLS will perform during nominal operation on Mars), and using 39-point lines (the maximum number of spots per sample foreseen to be analyzed by RLS). The resulting curves are represented in Figure 07, and correspond to the following equations:

$$proportion_{olivine}(20\ spots) = -0.007394 \cdot r_{oli}^2 + 1.742 \cdot r_{oli} - 0.5958; \quad R^2 = 0.9995$$
$$proportion_{serpentine}(20\ spots) = 0.007394 \cdot r_{serp}^2 - 1.742 \cdot r_{serp} + 100.6; \quad R^2 = 0.9993$$

$$proportion_{olivine}(39\ spots) = -0.007328 \cdot r_{oli}^2 + 1.736 \cdot r_{oli} - 0.5731; \quad R^2 = 0.9994$$
$$proportion_{serpentine}(39\ spots) = 0.007328 \cdot r_{serp}^2 - 1.736 \cdot r_{serp} + 100.6; \quad R^2 = 0.9994$$

We calculated the estimated uncertainty within 95% confidence bounds by calculating two-times the standard deviation of the estimated proportions for the different lines. This uncertainty, for 20-point lines is calculated as ±2σ = ±5.2% and for 39-point lines is ±2σ = ±3.0%. In other words, the higher the number of points per line, the better the quantification accuracy of the method.

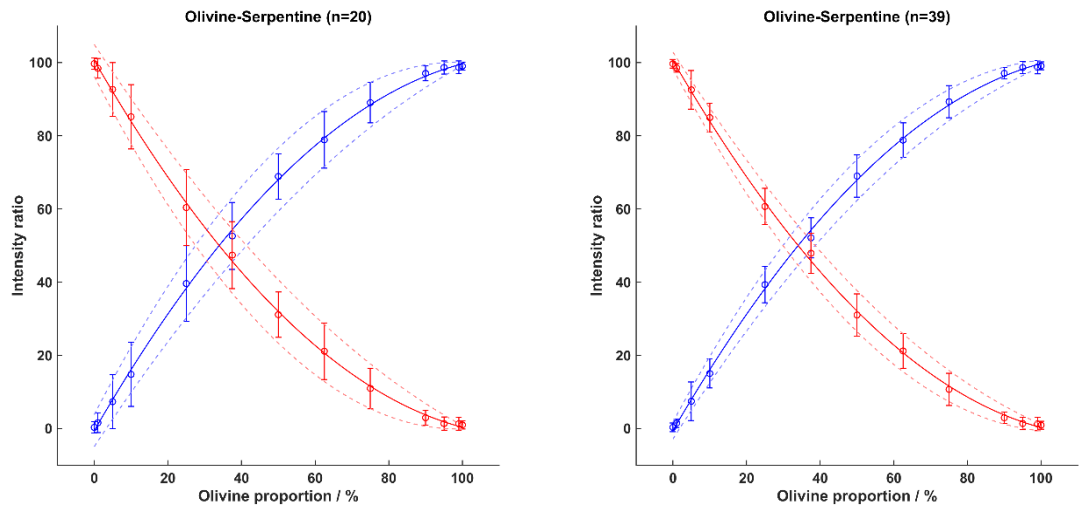

*Figure 07: Serpentine and olivine calibration curves for 20-point lines (left) and 39-point lines (right). Error bars show the quantification uncertainties of the method.*

As already explained, the RLS instrument adjusts the acquisition parameters to the sample under analysis. This way, it is possible to save time from the total operational sequence that can be used to analyze further points. This adaptive procedure is implemented in the rover and is responsible for the uncertainty on the total number of acquired spots per sample by the RLS instrument. The nominal case of 20 points per sample can be increased up to 39 points if the analysis of the first spots requires less time than allocated. As described in section 2.2.3 we calculated the estimated proportion uncertainty of the calibration curves for different numbers of points per line (from 1 to 39), to infer the semi-quantification capabilities of the RLS. Though the higher the number of points, the smaller the prediction uncertainty, Figure 08 shows that a semi-quantification of the mineral abundances would be possible when working within the foreseen operational parameters of RLS (between 20 and 39 spots).

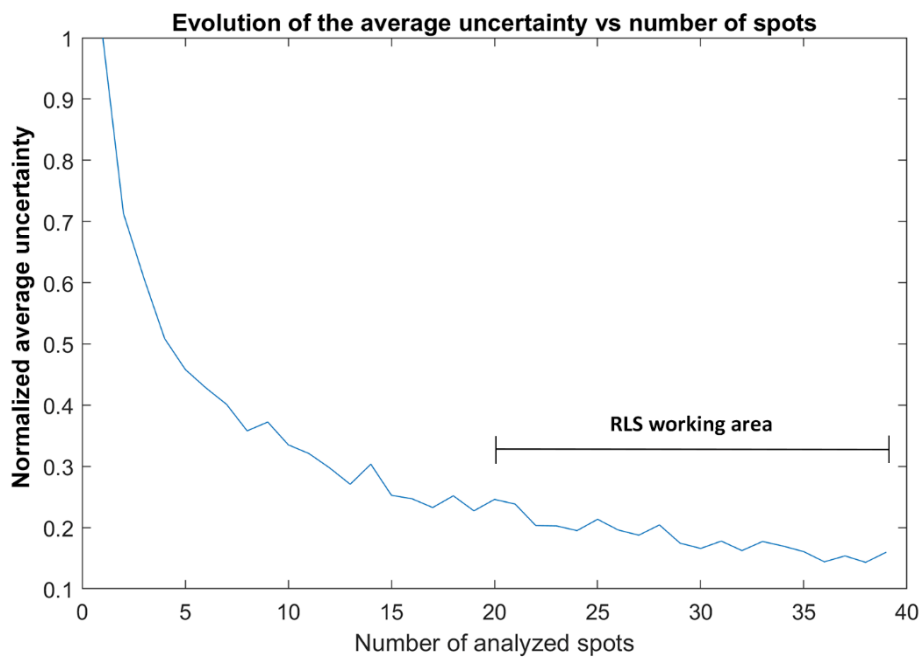

*Figure 08: Prediction uncertainty as function of the number of analyzed spots per line.*

**3.4 Estimation of serpentinization degree of Leka ophiolite samples (RLS ExoMars Simulator)**

By using the RLS ExoMars Simulator, we acquired five lines of 39 spots from each Leka ophiolite sample. In order to evaluate the abundance determination uncertainties, we also used subsets of 20 points per line to obtain the results with the 20-point calibration curve. Table 02 shows the quantification results obtained for the five lines for each Leka samples, when run through the 20 and 39-point calibration curves. Two values of uncertainty are provided. One is the uncertainty from the calibration curve. This value is obtained from the extrapolation of the uncertainty value calculated for the calibration curve at the detected proportion of the samples. This is the inherent uncertainty associated with the quantification method, i.e. the uncertainty with 95% confidence bounds on the calibration curves. This value depends on the concentration of the sample (as can be observed by the uncertainty lines shown in Figure 07). The second value of uncertainty is the deviation in the results between different lines of analysis on the same sample. On Mars, the RLS instrument will acquire one line of spectra instead of five. Our results show that the expected uncertainty in the quantification results due to the analysis of only one line is 4.0% or less, depending on the sample.

*Table 02: Comparison of semi-quantitative results extrapolated from the interpretation of RLS ExoMars Simulator data.*

| Sample | Line | Abundance estimation (39 point/line) | | | | Abundance estimation (20 points/line) | | | |
|---|---|---|---|---|---|---|---|---|---|
| | | Olivine | Serpentine | Uncert. (from cal. curve) | Std between lines | Olivine | Serpentine | Uncert. (from cal. curve) | Std between lines |
| LE16-0002 | 1 | 3 | 97 | 3 | 0 | 3 | 97 | 5 | 0 |
| | 2 | 2 | 98 | 3 | | 2 | 98 | 5 | |
| | 3 | 2 | 98 | 3 | | 2 | 98 | 5 | |
| | 4 | 2 | 98 | 3 | | 3 | 97 | 5 | |
| | 5 | 2 | 98 | 3 | | 3 | 97 | 5 | |
| | 1 | 42 | 58 | 6 | 2 | 42 | 58 | 9 | 3 |

| Sample | | | | | | | | | |
|---|---|---|---|---|---|---|---|---|---|
| LE16-0004 | 2 | 40 | 60 | 6 | | 42 | 58 | 9 | |
| | 3 | 37 | 63 | 6 | | 36 | 64 | 9 | |
| | 4 | 39 | 61 | 6 | | 37 | 63 | 9 | |
| | 5 | 39 | 61 | 6 | | 40 | 60 | 9 | |
| LE16-0005 | 1 | 19 | 81 | 5 | 3 | 19 | 81 | 7 | 4 |
| | 2 | 17 | 83 | 4 | | 13 | 87 | 7 | |
| | 3 | 20 | 80 | 5 | | 17 | 83 | 7 | |
| | 4 | 24 | 76 | 5 | | 25 | 75 | 8 | |
| | 5 | 24 | 76 | 5 | | 19 | 81 | 7 | |
| LE16-0010 | 1 | 13 | 87 | 4 | 2 | 13 | 87 | 7 | 2 |
| | 2 | 11 | 89 | 4 | | 12 | 88 | 6 | |
| | 3 | 10 | 90 | 4 | | 9 | 91 | 6 | |
| | 4 | 14 | 86 | 4 | | 13 | 87 | 7 | |
| | 5 | 10 | 90 | 4 | | 11 | 89 | 6 | |
| LE16-0017 | 1 | 25 | 75 | 5 | 3 | 26 | 74 | 8 | 4 |
| | 2 | 21 | 79 | 5 | | 19 | 81 | 7 | |
| | 3 | 26 | 74 | 5 | | 30 | 70 | 8 | |
| | 4 | 23 | 77 | 5 | | 20 | 80 | 8 | |
| | 5 | 20 | 80 | 5 | | 22 | 78 | 8 | |

## 4 Discussion

As mentioned in the introduction section, the ALD of the *Rosalind Franklin* rover will be equipped with three main analytical systems. MicrOmega will examine crushed materials collected from the Martian subsurface to characterize their mineralogical composition at grain-size level. NIR data will be combined to point by point Raman analysis (RLS system) in two different operative modes: cooperative and automatic. In cooperative mode, the RLS instrument will analyze interesting spots of the sample based on the MicrOmega analysis, while RLS working in automatic mode will analyze between 20 and 39 spots of the sample along a line. Irrespective of the acquisition mode, the combination of Raman and NIR data will be of key importance in the selection of the potential biomarkers-bearing samples to be analyzed by MOMA (Vago et al., 2017). Having this in mind, our results (section 3.1) prove that analytical procedures based on the combination of NIR and Raman analysis ensure a reliable mineralogical characterization of Martian analogues. Indeed, as shown on Table 03, the major mineral phases composing Leka samples (olivine, serpentine and pyroxene), were effectively detected by both spectroscopic techniques. On one hand, minor compounds such as Fe-oxides and Mg-hydroxides were not detected by NIR. However, unlike the instrument used in this work, MicrOmega will acquire a near-infrared spectrum from each of the 64000 pixels (size of 20 µm$^2$) composing its field of view (5 mm$^2$). Thus, from the detailed analysis of MicrOmega data sets, the mineralogical heterogeneities of Martian subsurface samples will be more easily detected. NIR data were also used to extrapolate inferences about the alteration degree of powdered samples: spectroscopic results fit with XRD semi-quantitative estimations, identifying LE16-0002 as the most altered sample and LE16-0005 and LE16-0017 as the less altered ones. Considering the field-of-view of analysis of MicrOmega is 5mm$^2$ (25 times larger than the spectrometer used in this work), it can be assumed that NIR data gathered on Mars will provide the possibility of performing more representative semi-quantitative studies.

As described elsewhere, when ALD instruments operate in cooperative mode, MicrOmega multispectral images will be employed by the ExoMars science team to determine the spot of interest to be deeply analyzed by RLS (Vago et al., 2017). The results we presented in sections 3.1.3 and 3.2 shed light on the advantages offered by this analytical approach. Through the interpretation of high-quality spectra collected by both microRaman and RLS ExoMars Simulator systems we obtained a more detailed characterization of major compounds: the Fo/Fa ratio of olivine grains was clearly determined while serpentine polytypes (antigorite, lizardite) were effectively distinguished. Furthermore, numerous minor compounds were also identified (such as hematite and brucite), some of which were not detected by XRD (e.g. ilmenite on sample LE16-0005 and calcite on sample LE16-0017). These results underline one of the greatest advantages ensured by the use of Raman spectrometers: the laser beam can be focused on micrometric grains (in the case of the RLS soon operating on Mars, the spot of analysis is about 50 µm), which greatly facilitates the detection of minor mineral phases. This is an aspect of high relevance for the future ExoMars mission since minor phases bears relevant information that could help reconstructing formational and alteration processes occurred or still occurring in the subsurface of Mars.

As indicated by Raman data comparison, some minor/trace compounds were more easily detected by the microRaman system than by the RLS ExoMars Simulator. This difference is justified by the fact that we used microRaman by visually selecting the spots of interest, while they were randomly chosen during RLS ExoMars Simulator analyses (automatic mode). In addition, it should be highlighted that the Raman resonance phenomenon between the excitation source of the microRaman (633nm, red laser) and the electronic transitions of ilmenite, magnetite and hematite enhanced the vibrational modes of these minerals, facilitating their detection. This aspect proves that the Raman detection of certain compounds strongly depends on the wavelength of the laser employed for their excitation. Considering the primary objective of the ExoMars mission is the detection of biomarkers, the choice of equipping the RLS system with a 532nm excitation laser rely on the Raman efficiency and the detector's spectral response ensured by this wavelength. Indeed, as the RLS team justify elsewhere (Rull et al., 2017), these two aspects make the 532nm laser the best option to enhance the detection of Raman peaks in the spectral range above 2800 cm$^{-1}$, where most of the vibrational features of CH- and OH- functional groups are observed. The spectra represented in Figure 06 support the decision of selecting a green laser source: Raman bands generated by the vibration of serpentine-interlayered water molecules were clearly detected by the RLS ExoMars Simulator. These spectra, collected by emulating the automatic operating mode of the RLS on Mars, also demonstrate that OH- vibrational bands vary according to the serpentine polytype under analysis (main peaks at 3674 and 3694 cm$^{-1}$ from antigorite grains, main peaks at 3640, 3689 and 3706 cm$^{-1}$ from lizardite grains).

*Table 03: qualitative result comparison.*

| LE16-0002 | | LE16-0004 | | LE16-0005 | | LE16-0010 | | LE16-0017 | |
|---|---|---|---|---|---|---|---|---|---|
| Serpentine | ▨ | Serpentine | ▨ | Serpentine | ▨ | Serpentine | ▨ | Serpentine | ▨ |
| Pyroxene | ☐ | Olivine | ▨ | Olivine | ▨ | Olivine | ▨ | Olivine | ▨ |
| Fe-oxides | ▨ | Pyroxene | ▨ | Pyroxene | ▨ | Mg-hydroxide | ▨ | Fe-oxides | ▨ |
| | | Mg-hydroxide | ▨ | Mg-hydroxide | ▨ | Fe-oxides | ▨ | Pyroxene | ▨ |
| | | Fe-oxides | ▨ | Fe-oxides | ▨ | Pyroxene | ▨ | Calcite | ▨ |
| | | | | Ilmenite | ▨ | Unknown phyllosilicate | ▨ | | |

Legend: ☐ XRD | ▨ NIR | ▨ microRaman | ▨ RLS ExoMars Simulator

For both automatic and cooperative operating modes, the RLS has been programmed to acquire multiple consecutive analysis over powdered samples, ranging a number of spots between 20 and 39. Depending on mineralogical heterogeneity, the set of Raman data gathered from each Martian target could be therefore used to estimate the content of the detected mineral phases. Thanks to the results summarized in sections 3.3 and 3.4 and obtained by simulating the automatic operating mode of the RLS on Mars, we proved that Raman data can be effectively used for the estimation of the content of the detected mineral phases. We employed the ExoMars Simulator to perform 39 automatic analysis on each olivine/serpentine mixture and we used the obtained spectra to correlate peak intensities with mineral contents. Figure 08 shows that the uncertainty of semi-quantitative results decreases significantly as the number of analyzed spots increases, reaching fairly low values in the range between 20 and 39 spectra. We then used the set of Raman data to generate calibration curves that, as detailed in section 3.3, ensured correlation coefficients above 0.9993 and uncertainty values between 3.0 and 5.2%.

To validate the calibration curves, we applied the equations presented in section 3.3 to the Raman spectra gathered from LOC analogues. The semi-quantification values were finally compared to those provided by XRD. As displayed in Figure 09, our comparison showed a good correlation between the two techniques, which confirms the possibility of using Raman data in automatic mode to determine the serpentinization degree of olivine-rich samples.

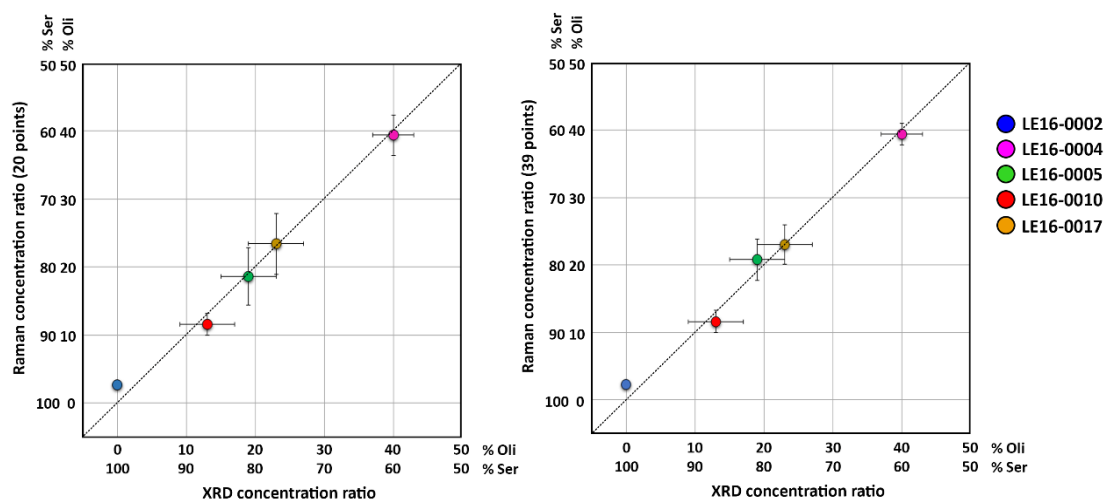

*Figure 09: RLS vs XRD semi-quantification for 20 and 39-line points. The error bars show the quantification uncertainties for both techniques.*

Among the 5 samples evaluated in this study, LE16-0002 is the one providing the most discordant results: we detected minor amounts of olivine from RLS ExoMars Simulator data, while XRD data seggested the total serpentinization of this mineral. This difference further highlights that, as previously explained, multiple Raman analysis often enables the identification of mineral phases whose contents are below the detection limits of XRD systems or other macro techniques.

## Conclusions

The comparison between diffractometric and spectroscopic data gathered from laboratory systems proves that combined Raman and NIR analysis ensures a reliable identification of both primary minerals (i.e. olivine, pyroxene, ilmenite and feldspar) and alteration products (i.e. serpentine, calcite, brucite and hematite) composing LOC analogues. By carrying out a refined data interpretation, we proved that Raman data also serves to estimate the Fo/Fa ratio of olivine grains (between $Fo_{87}Fa_{13}$ and $Fo_{92}Fa_{08}$) as well as to distinguish the presence of different serpentine polytypes (lizardite and antigorite). The same analogues were then studied by means of the RLS ExoMars Simulator, which was employed by following the automatic operating mode of the RLS instrument soon operating on Mars onboard the *Rosalyn Franklin* rover. The obtained data are qualitatively comparable to those provided by the laboratory Raman system. Our results confirm that the RLS will play a key role in the mineralogical characterization of Martian samples, especially as regards the identification of minor mineral phases, both in cooperative and automatic acquisition modes. We have shown how the overall qualitative interpretation of data of minor compounds (<1%) could be improved by the cooperative operation of the different techniques, with RLS analyzing interesting spots detected by MicrOmega. However, as demonstrated by the analytical data summarized in section 3.3, the RLS ExoMars simulator data obtained in automatic mode can not only detect a good number of minor compounds that might be missed by IR analysis, but also reliably estimate the olivine/serpentine content ratio of powdered rocks, being the obtained correlation coefficients ($R^2$) equal to 0.9993 (20 points) and 0.9995 (39 points) with an estimated uncertainty between ±3.0% and ±5.2% with a confidence interval of 95%. Data sets gathered from the Raman study of standard mixtures also proves that the RLS ExoMars Simulator can detect both olivine and serpentine down to a concentration of 1%. Furthermore, we also demonstrate that the RLS automatic operating mode allows the semi-quantification of the mineral abundance of binary mixtures. We applied olivine/serpentine calibration curves to the set of Raman data gathered from LOC analogues. As suggested by the comparison with XRD semi-quantification data, Raman spectroscopy can effectively determine the degree of serpentinization of olivine-rich samples based on the spectra acquired in RLS automatic operating mode (at least 20 points randomly acquired on a powdered sample). Taking into account the importance of olivine serpentinization related to microorganisms proliferation and biomarkers accumulation, the results presented in this work underline the key role the RLS system will play in the selection of the optimal scientific target rocks to be studied on Mars for the detection of life tracers. In this sense, our research must be interpreted as part of a broader research line aiming to define the experimental tools necessary for the optimal interpretation of RLS data from Mars.

**Acknowledgements**

This work is financed through the European Research Council in the H2020- COMPET-2015 programme (grant 687302) and the Ministry of Economy and Competitiveness (MINECO, grant ESP2017-87690-C3-1-R).This work is financed through the European Research Council in the H2020- COMPET-2015 programme (grant 687302) and the Ministry of Economy and Competitiveness (MINECO, grant ESP2017-87690-C3-1-R).